\documentclass[prl,twocolumn,floatfix,superscriptaddress]{revtex4-1}
\usepackage{dcolumn,amsmath}
\usepackage{graphicx}
\usepackage{bm}
\usepackage{hyperref}
\usepackage{xcolor}

\usepackage[utf8]{inputenc}
\usepackage[T1]{fontenc}

\newcommand{\Eeff}{\ensuremath{E_{\rm eff}}}
\newcommand{\eEDM}{{\em e}EDM}

\usepackage{tabularx}
\usepackage{braket}

\begin{document}

\title{Laser-coolable AcOH$^+$ ion for $\mathcal{CP}$-violation searches}

\author{Alexander V. Oleynichenko}
\email{oleynichenko\_av@pnpi.nrcki.ru \\ alexvoleynichenko@gmail.com}
\affiliation{Petersburg Nuclear Physics Institute named by B.P.\ Konstantinov of National Research Center ``Kurchatov Institute'' (NRC ``Kurchatov Institute'' - PNPI), 1 Orlova roscha, Gatchina, 188300 Leningrad region, Russia}
\homepage{http://www.qchem.pnpi.spb.ru}

\author{Leonid V.\ Skripnikov}
\email{skripnikov\_lv@pnpi.nrcki.ru,\\ leonidos239@gmail.com}
\affiliation{Petersburg Nuclear Physics Institute named by B.P.\ Konstantinov of National Research Center ``Kurchatov Institute'' (NRC ``Kurchatov Institute'' - PNPI), 1 Orlova roscha, Gatchina, 188300 Leningrad region, Russia}
\affiliation{Saint Petersburg State University, 7/9 Universitetskaya nab., St. Petersburg, 199034 Russia}

\author{Andr\'ei V. Zaitsevskii}
\affiliation{Petersburg Nuclear Physics Institute named by B.P.\ Konstantinov of National Research Center ``Kurchatov Institute'' (NRC ``Kurchatov Institute'' - PNPI), 1 Orlova roscha, Gatchina, 188300 Leningrad region, Russia}
\affiliation{Department of Chemistry, M.V. Lomonosov Moscow State University, Leninskie gory 1/3, Moscow, 119991~Russia}

\author{Victor V.\ Flambaum}
\affiliation{School of Physics, The University of New South Wales, Sydney NSW 2052, Australia}
\affiliation{Johannes Gutenberg-Universit\"at Mainz, 55099 Mainz, Germany}

\begin{abstract} 
The AcOH${}^+$ molecular ion is identified as a prospective system to search for $\mathcal{CP}$-violation effects. According to our study AcOH${}^+$ belongs to the class of laser-coolable polyatomic molecular cations implying the large coherence time in the experiments to study symmetry violating effects of fundamental interactions. We perform both nuclear and high level relativistic coupled cluster electronic structure calculations to express experimentally measurable $\mathcal{T}$,$\mathcal{P}$-violating energy shift in terms of fundamental quantities such as the nuclear magnetic quadrupole moment (MQM), electron electric dipole moment ($e$EDM) and dimensionless scalar-pseudoscalar nuclear-electron interaction constant. We further express nuclear MQM in terms of the strength constants of $\mathcal{CP}$-violating nuclear forces: quantum chromodynamics vacuum angle $\bar{\theta}$ and quark chromo-EDMs. The equilibrium geometry of AcOH${}^+$ in the ground and the four lowest excited electronic states was found to be linear. The calculated Franck-Condon factors and transition dipole moments indicate that the laser cooling using optical cycle involving the first excited state is possible for the trapped AcOH${}^+$ ions with the Doppler limit estimated to be~$\sim 4$~nK. The lifetime of the (0,1$^1$,0) excited vibrational state considered as a working one for MQM and $e$EDM search experiments is estimated to be $\sim 0.4$ sec.
\end{abstract}

\maketitle

\section{Introduction}

Violation of the time-reversal ($\mathcal{T}$) and spatial parity ($\mathcal{P}$) symmetries of fundamental interactions is one of the most intriguing problems of modern physics. According to the $\mathcal{CPT}$ theorem, violation of the $\mathcal{T}$-symmetry implies violation of the $\mathcal{CP}$ symmetry, where $\mathcal{C}$ is the charge conjugation. $\mathcal{CP}$-violation has been discovered in the decays of neutral kaons more than half a century ago. According to Sakharov $\mathcal{CP}$-violation is a key ingredient in understanding the baryogenesis problem \cite{sakharov1967violation} and is important for cosmology and astrophysics. However, the known level of $\mathcal{CP}$-violation in the standard model cannot explain the observed ratio of the number of protons to number of photons in the Universe related to the predominance of the matter over antimatter.
At present $\mathcal{CP}$-violation is known for the weak interactions, but not for the   quantum chromodynamics (QCD) sector. The very small magnitude of the CP violation in QCD is hard to explain within the standard model. This  is known as a strong $\mathcal{CP}$-problem. Therefore, numerous experiments are devoted to searching for $\mathcal{CP}$-violation and it is expected that they will be able to shed light on these problems.

In the second half of the 20th century, it was realized that heavy atoms are very good  systems to search for $\mathcal{T},\mathcal{P}$-violation effects.  Moreover, diatomic molecules and ions containing heavy atoms can be much more sensitive to such effect due to the closeness of opposite parity energy levels~\cite{Flambaum:76,Labz78,SF78}. There is a variety of possible manifestations of $\mathcal{T}, \mathcal{P}$-violation effects which can be treated. $\mathcal{CP}$-violation forces inside the nucleus can lead to the nonzero value of the nuclear magnetic quadrupole moment (MQM). This moment can interact with electrons of a molecules with open electronic shells and induce energy shift, which can be measured. MQM can be expressed in terms of the fundamental parameters of the interactions such as the quantum chromodynamics (QCD) parameter ${\bar \theta}$ and other hadronic $\mathcal{CP}$-violation parameters~\cite{FDK14,Flambaum:19a,Flambaum:2020a,Engel:2003,Dobaczewski:2018}. Thus, molecular systems can be sensitive to these parameters.
Besides, experiments with such heavy atom molecules can be very sensitive to the electron electric dipole moment (\eEDM) and scalar-pseudoscalar nucleus-electron interaction. Indeed the current best constraint on \eEDM, $|d_e|<1.1\cdot 10^{-29} e\cdot \textrm{cm}$, has been obtained on the $^{232}$ThO molecule by the ACME collaboration~\cite{ACME:18}.

Compounds of other 7th row elements of the periodic table are also very promising for considered molecular experiments. In particular due to their high nuclear charges resulting in large enhancement (see below) of $\mathcal{T}$,$\mathcal{P}$-violating effects. Another reason is a wide variety of known isotopes suitable for studying various phenomena. In addition to the aforementioned ThO molecule, similar experiments aimed at precision measurement of the electron EDM are now carried out using the ThF$^+$ cation~\cite{Gresh:2016}. Radium-containing molecules actively studied in the recent years are also considered as very promising candidates for such measurements \cite{Isaev:2010,Kudashov:13,Isaev:2013,Borschevsky:13, Kudashov:14, Sudip:2016b, isaev2017laser, Petrov:2020, Osika:20,Skripnikov:2020e, Isaev:2021,Skripnikov:2021a, Yu:2021, Osika:21, Osika:22, zakharova2021rotating}. Atomic francium was proposed for $\mathcal{T}, \mathcal{P}$-violation studies by the FrPNC collaboration at TRIUMF \cite{FrPNC:2015,shitara2021cp}. Theoretical proposals consider even the lawrencium diatomics for which the \eEDM~enhancement factor and the constant of the electron-nucleus scalar-pseudoscalar interaction constant were calculated recently \cite{Mitra:2021}. In this regard it seems especially surprising that quite small attention was paid to actinium compounds. Meanwhile at the moment 29 isotopes of actinium are known, 13 of them (${}^{222m}$Ac - ${}^{234}$Ac) having lifetimes in the range from 40 seconds (${}^{234}$Ac) to 21.77 years (${}^{227}$Ac) \cite{TableOfIsotopes}. The ${}^{224}$Ac isotope (2.78 hours) possess nuclear spin $I=0$ and thus seems to be perfect for \eEDM~searches. All other isotopes have non-zero nuclear spins, and for isotopes with the neutron number $130 < N < 142$ significant octupole deformations are expected from both theoretical and experimental data \cite{Verstraelen:2019}, which can result in existence of nuclear levels of opposite parity and hence strongly enhanced $\mathcal{CP}$-violating nuclear moments \cite{Flambaum:2020a}. This gives opportunities for the $\mathcal{CP}$-violation searches using actinium compounds such as  AcF, AcO$^+$ and AcN molecules~\cite{Flambaum:19a,Skripnikov:2020c}. In the recent report \cite{Kyuberis:2021} the values of the effective electric field required to interpret the \eEDM~experiments were calculated for the AcOH${}^+$ and AcOCH$_3^+$ molecular ions.

Recently it has been suggested to perform experiments to search for the $\mathcal{T}$,$\mathcal{P}$-violating effects with linear triatomic molecules and ions~\cite{kozyryev2017precision,Denis:2019,Maison:2019b,Maison:20a,Zakharova:RaOH:2021,Zakharova:2021,Maison:2021,maison2021axion}. With such systems it is expected to probe high-energy physics beyond the standard model in the PeV regime~\cite{kozyryev2017precision}. Such systems as YbOH in the ground electronic state $^2\Sigma_{1/2}$ can benefit from easy polarization by weak external electric fields~\cite{kozyryev2017precision} in comparison with diatomic analog YbF. This feature of triatomic quasilinear molecules is due to the $l$-doubling effect resulting in the existence of the two closely spaced levels of opposite parity. At the same time YbOH (as YbF) can be laser cooled to very low temperatures. Experiments with such molecules thus will benefit from the very large coherence time since the uncertainty of the measurement of $\mathcal{T}, \mathcal{P}$-violating effects is inversely proportional to this time. 
Indeed, in recent years significant progress in laser cooling of polyatomic molecules has been made. The successful one-dimensional direct laser cooling of the SrOH \cite{Kozyryev2017}, CaOH \cite{Baum2020}, YbOH \cite{Augenbraun2020} and CaOCH${}_3$ \cite{Mitra2020} molecules has been already demonstrated. Direct laser cooling was also predicted to be feasible for RaOH~\cite{isaev2017laser}, YbOCH$_3$~\cite{Augenbraun:2021}. Recently, there was a proposal to search for $\mathcal{T}, \mathcal{P}$-violating effects on the three-atomic molecular ion LuOH$^+$ which can be sympathetically cooled~\cite{Maison:20a}. In the present paper we study a system of another type, a three-atomic cation which can potentially be laser cooled. From the theoretical point of view, molecular ions must satisfy actually the same requirement as molecules in order to be laser cooled, i.e. nearly diagonal Franck-Condon matrix for the working electronic transition. To the moment only some diatomic ions have been concerned as a candidates for the cooling procedure, namely, BH${}^+$ \cite{Nguyen2011}, AlH${}^+$ \cite{Nguyen2011,Huang2020}, SiO${}^{+}$ \cite{Qin2020,Stollenwerk2020}, TeH${}^+$ \cite{Stollenwerk2018}, OH${}^{-}$ \cite{Wan2017}, C${}_2^{-}$ \cite{Yzombard2015}, TlF${}^+$~\cite{Elmoussaoui2021} and some other diatomic cations \cite{Ivanov2020}. To the best of our knowledge none of these ions has been laser cooled yet.

The paper is organized in the following way. Firstly we describe details of the electronic structure models used. Secondly we present potential energy surfaces for the ground and several electronic states of AcOH$^+$, dipole moments of transitions between them, Franck-Condon factors and discuss the relevance of the molecule for direct laser cooling. Then we express nuclear MQM of ${}^{225}$Ac and ${}^{227}$Ac isotopes in terms of the strength constants of $\mathcal{CP}$-violating nuclear forces, QCD vacuum angle $\bar{\theta}$, and quark chromo-EDMs $\tilde{d}_u$ and $\tilde{d}_d$. Finally, we discuss some technical questions concerning the relevance of the AcOH${}^+$ ion for $\mathcal{T}, \mathcal{P}$-violation search experiments and report the precise calculations of the molecular $\mathcal{T}, \mathcal{P}$-violating parameters required to interpret experimental data.

\section{Computational details}\label{sec:comput}

\subsection{Excited state calculations}
Qualitatively, the ground and low-lying electronic states of AcOH$^+$ can be described as having one unpaired electron above the closed-shell AcOH$^{2+}$ configuration. In other words, these states can be described within
the one-particle $0h1p$ Fock space sector, assuming the ground closed-shell state of AcOH$^{2+}$ as a Fermi vacuum.
Thus the relativistic Fock space coupled cluster method with single and double excitations (FS-RCCSD) \cite{Kaldor:1991,Visscher:2001,Eliav:Review:2016} was applied to obtain excitation energies. This method had proven itself as a highly precise and reliable tool for quite similar systems, e.g. RaOH \cite{isaev2017laser}, RaF \cite{Kudashov:14,Zaitsevskii:RaF:2021,Osika:22}, YbOH \cite{Denis:2019,Maison:2019b}. In order to make sure that there are no low-lying charge-transfer electronic states, additional FS-RCCSD calculations starting from the neutral AcOH molecule ground state as an alternative Fermi vacuum were conducted in the one-hole $1h0p$ Fock space sector. The lowest charge transfer-type state was found to lie above 50000 cm$^{-1}$ and thus does not deteriorate somehow the picture of low-lying electronic states. Within the paper only the five lowest-lying states, namely, the (1-2)1/2, (1-2)3/2 and (1)5/2 were considered. Electronic transition dipole moment functions were evaluated using the FS-RCCSD method combined with the finite-field technique \cite{Zaitsevskii:1998,zaitsevskii2018electronic}.

Although the FS-RCCSD method reproduces excitation energies fairly well, the neglect of higher excitations as well as the basis set superposition errors (BSSE) introduce systematical (and quite uniform) errors into the dependencies of all electronic state energies on molecular geometry parameters. To reduce these errors, the technique adopted from \cite{Pazyuk:2015,Isaev:2021} was employed: the ground state potential energy surface was computed within the single reference relativistic CCSD(T) method, taking into account triple excitations in a perturbative manner, with the counterpoise correction for BSSE. Potential energy surfaces for excited states were obtained by combining the resulting ground state surface with the FS-RCCSD excitation energies as functions of the internuclear distances and the valence angle.

Accurate {\it ab initio} modeling requires an adequate description of relativistic effects, including those beyond the conventional Dirac–Coulomb Hamiltonian. The effect of the Breit interaction on the excitation energies of atomic Ac and its ions was demonstrated to be non-negligible in \cite{Eliav:1998}. For this reason the relativistic model used in the present paper was defined by accurate two-component relativistic pseudopotentials (RPPs) derived from the atomic Dirac-Fock-Breit calculations and thus implicitly incorporating the bulk of Breit interaction effects. These RPPs correspond to the valence part of the generalized relativistic pseudopotentials, developed in Refs.~\cite{Mosyagin:2010,Mosyagin:2016,Mosyagin:2020,ECP_website}. The small-core RPP replacing the shells with principal quantum number $n \le 4$ \cite{Mosyagin:2016} was used for Ac, and the recently developed ``empty-core'' RPP leaving all 8 electrons for the explicit treatment \cite{Mosyagin:2020} was used for O. The pseudopotential approximation also allows one to use large and flexible contracted Gaussian basis sets for the explicit description of outer core and valence electrons. The basis set for Ac comprised primitive (15$s$14$p$10$d$8$f$) Gaussians loosely based on the exponents from the ANO-RCC set \cite{Roos:2005} and the contracted (6$g$5$h$4$i$)/[4$g$3$h$2$i$] part obtained as averaged natural atomic orbitals in scalar-relativistic CCSD calculations~\cite{Skripnikov:13a} of low-lying electronic states of Ac and Ac$^{2+}$. The aug-cc-pVQZ-DK and cc-pVQZ-DK basis sets \cite{Dunning:1989,DeJong:2001,Kendall:1992} were used for O and H atoms, respectively. In all coupled cluster calculations the $5s5p$ shells of Ac and $1s$ shell of O were not included in the correlation treatment.

Molecular integral transformation and CCSD(T) calculations were carried out using the {\sc dirac19} \cite{DIRAC19,Saue:2020} package. For FS-RCCSD calculations the {\sc exp-t} program system was employed \cite{Oleynichenko_EXPT,EXPT_website}. Lifetime of the excited vibrational (0,1,0) state was obtained using the dipole moment function estimated at the two-component Kramers-unrestricted relativistic density functional theory (DFT) level with the PBE0 functional \cite{Adamo:1999} using the code \cite{Wullen:2010}. To solve the one-dimensional vibrational problem and calculate Franck-Condon factors the {\sc vibrot} program \cite{VIBROT} was used. Spinors were visualized with the help of the {\sc vesta3} software \cite{VESTA3}.

\subsection{Calculations of the molecular $\mathcal{T}$,$\mathcal{P}$-violation parameters}

The nucleus with spin $I>1/2$ can possess nonzero nuclear magnetic quadrupole moment $M$. Nuclear calculations of $M$ for Ac isotopes are given in the next section. The interaction of MQM with electrons can be described by the following Hamiltonian~\cite{Ginges:04}:
\begin{equation} \label{HmqmOperator}
    H_{MQM} 
    =
    -\frac{M}{2I\left(2I-1\right)} T_{i,k} \cdot \frac{3}{2} \frac{\left[\boldsymbol{\alpha}\times \mathbf{r}\right]_i r_k}{r^5},
\end{equation}
where $T_{i,k}=I_i I_k+ I_k I_i -\frac{2}{3}I(I+1)\delta_{ik}$, $\boldsymbol{\alpha}$ are Dirac matrices. The electronic part of the Hamiltonian (\ref{HmqmOperator}) is characterized by the molecular constant $W_M$ given by~\cite{Sushkov:84,Kozlov:95}:
\begin{equation}
\label{WmME}
    W_M
    =
    \frac{3}{2 \Omega}
    \langle \Psi | \sum\limits_i 
    \left(
        \frac{\boldsymbol{\alpha}_i\times\mathbf{r}_i}
             {r_i^5}
    \right)_\zeta r_\zeta | \Psi \rangle,
\end{equation}
where sum index $i$ is over all the electrons and $\zeta$ is a projection on the molecular axis; $\Psi$ is the electronic wave function for the considered state of the molecule and
$\Omega=\langle\Psi|\bm{J}\cdot\bm{n}|\Psi\rangle$,
 $\bm{J}$ is the total electronic momentum. The resulting energy shift caused by the interaction of the nuclear MQM with electrons can be written as 
 \begin{equation}
\label{MQMshift}
 \delta E = C W_M M,
\end{equation}
 where parameter $C$ depends on the considered hyperfine sublevel of the molecule and the applied external electric field~\cite{Skripnikov:14a,Petrov:18b,Kurchavov:2020}.

Another source of $\mathcal{T}$,$\mathcal{P}$-violation effects in molecules is the electron EDM. Within the Dirac-Coulomb Hamiltonian the interaction induced by \eEDM~can be written in the following form~\cite{Lindroth:89}:
\begin{eqnarray}
  H_d^{{\rm eff}}= d_e\sum_j\frac{2i}{e\hbar}c\gamma^0_j\gamma_j^5\bm{p}_j^2,
 \label{Wd2}
\end{eqnarray}
where $d_e$ is the value of the \eEDM, index $j$ runs over electrons, $\bm{p}$ is the momentum operator for an electron and $\gamma^0$ and $\gamma^5=-i\gamma_0\gamma_1\gamma_2\gamma_3$ are the Dirac matrices, defined according to Ref. \cite{Khriplovich:91} 
\footnote{Note, that in the literature, there are two common definitions of $\gamma_5$ which differ by the sign.}. 
For a linear molecule this interaction can be characterized by the molecular constant $W_d$:
\begin{equation}
\label{matrelem}
W_d = \frac{1}{\Omega}
\langle \Psi|\frac{H_d}{d_e}|\Psi
\rangle.
\end{equation}
Here $\bm{n}$ is the unit vector along the molecular axis. In these designations effective electric field acting on the electron electric dipole moment is $E_{\rm eff}=W_d|\Omega|$.

The next considered source of $\mathcal{T}, \mathcal{P}$-violation is the nuclear spin-independent scalar-pseudoscalar nucleus-electron interaction. The interaction is given by the following Hamiltonian (see \cite{Ginges:04}, Eq.~(130)):
\begin{eqnarray}
  H_{T,P}=i\frac{G_F}{\sqrt{2}}Zk_{T,P}\sum_j \gamma^0_{j}\gamma^5_{j}\rho_N(\textbf{r}_j),
 \label{Htp}
\end{eqnarray}
where $G_F$ is the Fermi-coupling constant, $k_{T,P}$ is the dimensionless constant of the interaction, $\rho_N(\textbf{r})$ is the nuclear density normalized to unity and  $\mathbf{r}$ is the electron radius-vector with respect to the heavy atom nucleus under consideration.
%
This interaction is characterized by the molecular parameter $W_{T,P}$:
\begin{equation}
\label{WTP}
W_{T,P} = \frac{1}{\Omega}
\langle \Psi|\frac{H_{T,P}}{k_{T,P}}|\Psi
\rangle.
\end{equation}

Molecular parameters (\ref{WmME}), (\ref{matrelem}) and (\ref{WTP})  are required to interpret experimental data in terms of fundamental values of $M$, \eEDM~and $k_{T,P}$, respectively. To calculate them for the ground electronic state of AcOH${}^+$ we have used the following  scheme. The main correlation contribution to these parameters were calculated within the CCSD(T) approach using the Dirac-Coulomb Hamiltonian. All 97 electrons of AcOH$^+$ were included in this correlation calculation. The energy cutoff for virtual spinors has been set to 10000 Hartree. The inclusion of the high-energy virtual spinors is important for a correct description of the contribution of spin polarization and correlation effects of the inner-core electrons to the $\mathcal{T}, \mathcal{P}$-violating effects \cite{Skripnikov:17a,Skripnikov:15a}. In this calculation we have used the uncontracted Dyall's AETZ basis set for Ac~\cite{Dyall:07,Dyall:12} augmented by diffuse $s$-, $p$-, $d$- and $f$- type functions; $g$-type functions were reoptimized, thus the basis set for Ac comprised 37$s$-, 32$p$-, 21$d$-, 14$f$-, 8$g$- and 1$h$-type functions. For light atoms the aug-cc-pVTZ-DK basis set \cite{Dunning:1989,DeJong:2001,Kendall:1992} has been used.  In order to account for the basis set size correction we have extended basis set on Ac up to $(40s37p25d19f12g8h5i)$. This basis set corresponds to the uncontracted Dyall's AEQZ basis set for Ac~\cite{Dyall:07,Dyall:12} augmented by diffuse $s$-, $p$-, $d$- and $f$-type functions and partly reoptimized $g$-, $h$- and $i$-type functions. This basis set extension correction has been calculated within the CCSD(T) approach using the Dirac-Coulomb Hamiltonian. Inner-core electrons of Ac ($1s-3d$) were excluded from the correlation treatment and the virtual energy cutoff was set to 300 hartree in these calculations. 
We have also calculated contribution of the extended number of high harmonics up to $(15g15h15i)$. For this the two-step approach based on the generalized relativistic effective core potential theory~\cite{Titov:06amin, Petrov:02,Skripnikov:11a,Skripnikov:15b,Titov:96b,Skripnikov:16a} has been used. We employed the scalar-relativistic part of this operator. Therefore, it was possible to use very efficient scalar-relativistic code {\sc cfour}~\cite{CFOUR} and treat very large basis sets.
We have also estimated contribution of the Gaunt interelecton interaction as a difference between the values obtained at the Dirac-Hartree-Fock level within the Dirac-Coulomb-Gaunt and Dirac-Coulomb Hamiltonians.
Calculation of the property integrals were performed within the code developed in Refs.~\cite{Skripnikov:16b,Skripnikov:17b}.

\section{Results and discussion}

\subsection{Potential energy surfaces, Franck-Condon factors and relevance to laser cooling}

\begin{figure*}
\center
\includegraphics[width=\textwidth]{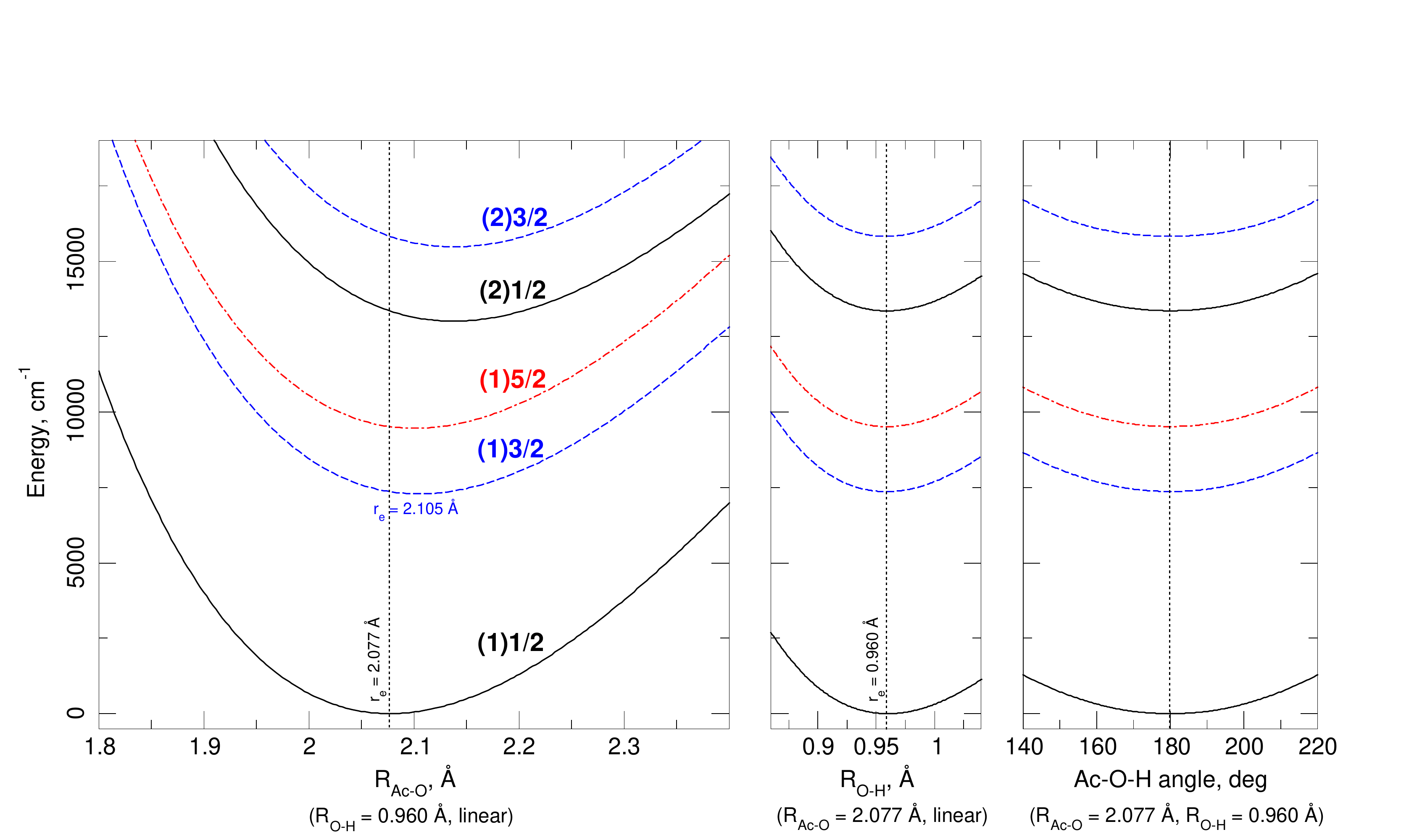}
\caption{
Potential energies of the ground (1)1/2 and excited states of the AcOH$^+$ ion as functions of R$_{\text{Ac-O}}$ (left), R$_{\text{O-H}}$ (center) and the valence Ac-O-H angle (right) with other parameters fixed at the equilibrium values for the ground state. Energies are given with respect to the ground state equilibrium point.
}\label{fig:pes}
\end{figure*}

Equilibrium geometries of AcOH$^+$ in its low-lying electronic states were optimized using the combined coupled cluster method described in the previous section.  AcOH$^+$ is predicted to be linear in all the electronic states considered; equilibrium bond distances and term energies are listed in Table~\ref{tab:equilibrium}. Several cross sections of four-dimensional potential energy surfaces are given in Fig.~\ref{fig:pes}. It should be emphasized that the $R_{\text{O-H}}$ bond distance actually remains unchanged upon electronic excitation. Vibrational frequencies were estimated using the assumption of valence forces (see \cite{HerzbergVol2} and references therein) and are also listed in Table~\ref{tab:equilibrium}. It can be seen that for the (1)3/2 and (1)5/2 electronic states the optimal geometries and frequencies closely resemble those of the ground state, thus leaving hope that the closed optical loop can exist for the $E1$-allowed (1)1/2 -- (1)3/2 transition. For the other two excited states (2)1/2 and (2)3/2 the geometry displacements are too large to keep the Franck-Condon matrix to be diagonal, and thus these states will not be discussed further.

\begin{table*}
\center
\caption{Calculated molecular constants of AcOH$^+$ in its low-lying electronic states and compositions of model wavefunctions in terms of their scalar relativistic counterparts at $r_e$((1)1/2) = 2.077~\AA. Frequencies are calculated for the ${}^{227}$AcOH$^+$ isotopologue. Compositions given in the last column were estimated using the technique described in \cite{Zaitsevskii:2017b}.}\label{tab:equilibrium}

\begin{tabular}{cccccccc}
\hline
\hline
State & $T_e$, cm${}^{-1}$ & \multicolumn{2}{c}{Equilibrium geometry} & \multicolumn{3}{c}{Normal modes} & Composition \\
 & & $r_e$(Ac-O) & $r_e$(O-H) & $\nu_1(\sigma^+)$ & $\nu_2(\pi)$ & $\nu_3(\sigma^+)$ & \\

\hline
(1)1/2 & 0     & 2.077 & 0.960 & 656 & 423 & 3868 & 100\% $^2\Sigma$ \\

(1)3/2 & 7294  & 2.105 & 0.958 & 626 & 424 & 3887 & 99\% $^2\Delta$ + 1\% $^2\Pi$ \\

(1)5/2 & 9469  & 2.101 & 0.959 & 629 & 425 & 3884 & 100\% $^2\Delta$ \\

(2)1/2 & 13020 & 2.138 & 0.958 & 603 & 413 & 3886 & 89\% $^2\Pi$ + 11\% $^2\Sigma$ \\

(2)3/2 & 15484 & 2.138 & 0.957 & 604 & 348 & 3898 & 99\% $^2\Pi$ + 1\% $^2\Delta$ \\
\hline
\hline
\end{tabular}
\end{table*}

\begin{figure}
\center
\includegraphics[width=0.5\textwidth]{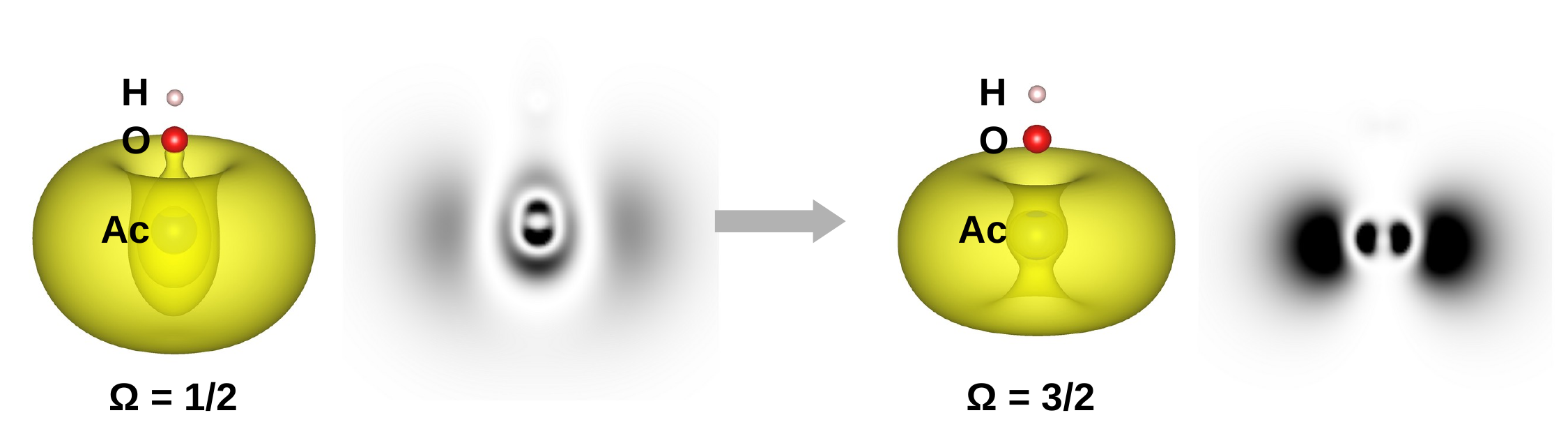}
\caption{Isosurfaces and cross sectional view of squared absolute values of approximate natural transition spinors (NTS) for the (1)1/2 $\rightarrow$ (1)3/2 transition in AcOH$^+$.}\label{fig:ntos}
\end{figure}

The fact that the (1)1/2 and (1)3/2 potential curves are nearly parallel to each other can be understood using simple one-electron picture \cite{IsaevHoekstra:2010,Ivanov:2019,Isaev:2021}. Fig.~\ref{fig:ntos} visualizes approximate natural transition spinors (NTS) for the (1)1/2-(1)3/2 transition derived from the model-space projections of many-electron wavefunctions; within the $0h1p$ Fock space sector, an electronic excitation can be represented by the only one pair of such approximate NTS. One can immediately see that one-particle densities for both hole and particle spinors are localized on the Ac atom. Thus the excitation essentially involves the non-bonding spinors which have most of their density outside the bonding region. Such an excitation should not result in any significant changes in bond lengths and force constants, thus leaving the matrix of Franck-Condon factors (FCFs) nearly diagonal.

To estimate vibrational branching ratios and Franck-Condon factors for the (1)1/2 -- (1)3/2 transition we neglect the anharmonic couplings between stretching and bending modes, so that the total FCF between the $(v'_1,v'_2,v'_3)$ and $(v''_1,v''_2,v''_3)$ vibrational states (where $v'_i$, $v''_i$ stand for vibrational quantum numbers) can be factorized \cite{Mengesha:2020}:
\begin{equation}
    \text{FCF} = |\braket{v'_1 v'_3|v''_1 v''_3}|^2 \times |\braket{v'_2|v''_2}|^2
\end{equation}
The latter factor is the one-dimensional bending mode FCF. For the AcOH$^+$ cation it is expected to be very close to unity since the corresponding sections of potential energy surfaces are perfectly parallel to each other. The overlap intergal over the $v_3$ stretch mode is expected to be nearly equal to unity and, which is more important, this factor will be the same for the $(v'_1,v'_2,0)$ and $(v''_1,v''_2,0)$ vibrational states analyzed here since this stretching mode is not excited in the states under consideration. Thus the total FCF can be approximated with the squared one-dimensional $\braket{v'_1|v''_1}$ overlap integral. The FCFs and branching ratios calculated within the Condon approximation are listed in Table~\ref{tab:fcfs}.

\begin{table}[h]
    \centering
    \caption{Estimated Franck-Condon factors (FCFs) and vibrational branching ratios (VBRs) for the (1)3/2$\rightarrow$(1)1/2 transitions.}
    \begin{tabular}{cccc}
    \hline
    \hline
    \multicolumn{2}{c}{Band} &  &  \\
    (1)3/2 & (1)1/2 & FCF & VBR, \% \\
    \hline
    \multicolumn{2}{c}{(0,0,0) $\rightarrow$ (0,0,0)} & 0.8892 & 91.6 \\
    \multicolumn{2}{c}{(0,0,0) $\rightarrow$ (1,0,0)} & 0.1058 & 8.1 \\
    \multicolumn{2}{c}{(0,0,0) $\rightarrow$ (2,0,0)} & 0.0049 & 0.3 \\
    \\
    \multicolumn{2}{c}{$\sum\text{FCF}$}  & 0.9999 \\
    \hline
    \hline
    \end{tabular}
    \label{tab:fcfs}
\end{table}

The estimated Franck-Condon factors and branching ratios closely resemble those obtained for the YbOH molecule within more precise discrete variable representation (DVR) method \cite{Mengesha:2020}. The sum of first three FCFs is equal to 0.9999, thus allowing one to conclude that the closed optical cycle can be implemented for the AcOH$^+$ molecular cation. Note that there are no electronic states lying between the ``working'' ones, and no additional decay pathways exist. The closed optical cycle will require at least three IR lasers, one at the working (0,0,0)~$\leftarrow$~(0,0,0) frequency and two repumping ones. Note that the (3,0,0)~$\leftarrow$~(0,0,0) and (0,$2^0$,0)~$\leftarrow$~(0,0,0) decay pathways are also allowed and in principle additional repumping lasers may be important to maximize the number of rescattered photons. A thorough analysis of FCFs would require accounting for the anharmonic couplings and deserves a separate detailed consideration in the future work.

\begin{figure}
\center
\includegraphics[width=0.45\textwidth]{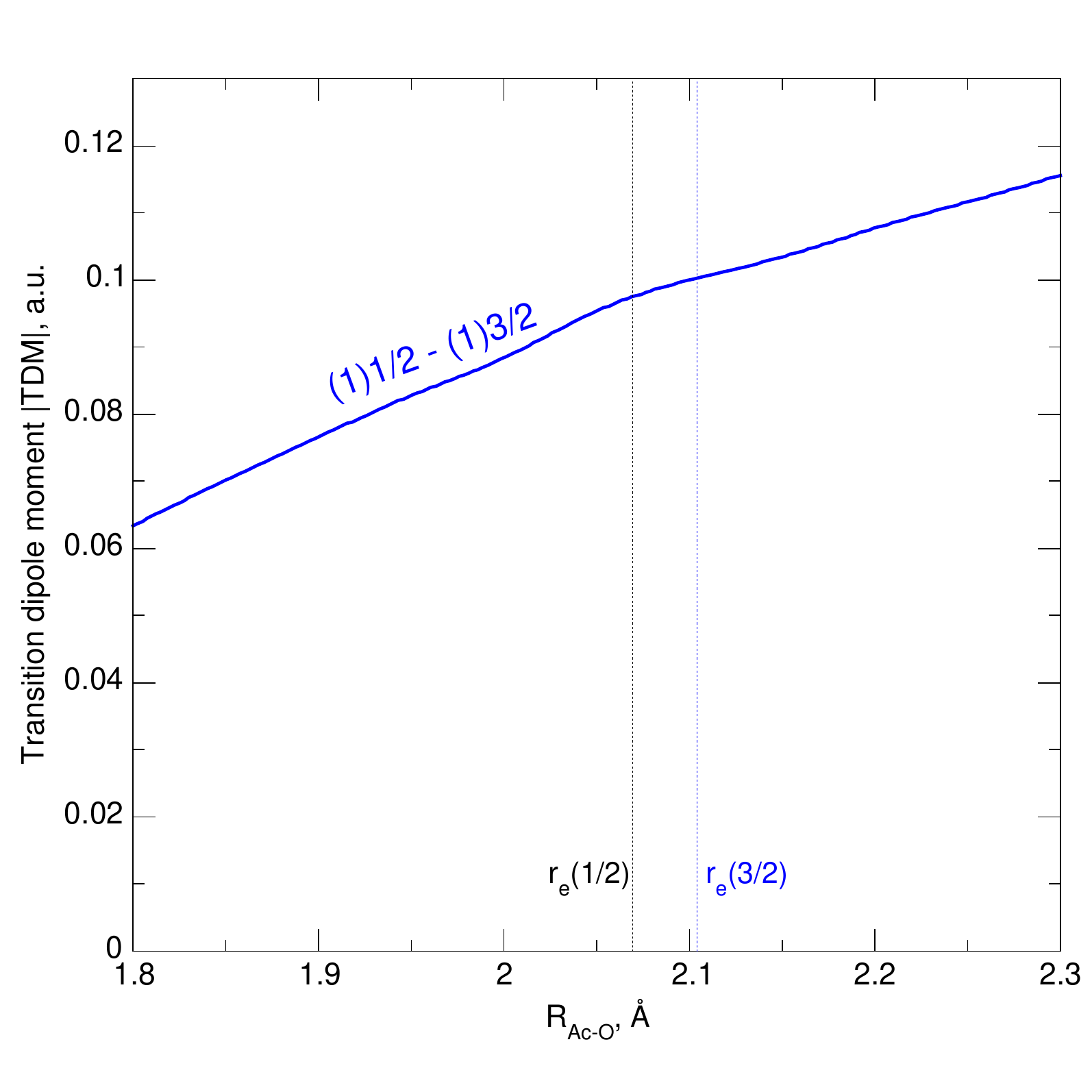}
\caption{
The absolute value of the (1)1/2 -- (1)3/2 transition dipole moment as a function of R$_{\text{Ac-O}}$, calculated at the FS-RCCSD level. Other parameters are fixed at the equilibrium values for the ground state (R$_{\text{O-H}}$ = 0.960 \AA, $\angle$(Ac-O-H) = 180$^{\circ}$).
}\label{fig:tdm}
\end{figure}

The last question which is to be discussed concerns the transition dipole moments and lifetime of the excited (1)3/2 state. It is well known that the excited state lifetime is of vital importance for laser cooling of neutral molecules. In order to achieve substantial slowing at least 10$^4$ photons are to be rescattered during the flight of the molecular beam through the setup. The shorter the lifetime, the more photons could be rescattered. For the polyatomic molecules previously cooled the excited state lifetimes do not exceed several dozens of nanoseconds \cite{Wall:2008,Mengesha:2020}. The situation for ions is quite different, since they can be trapped relatively easily. For example, the radio-frequency (rf) trap was recently used for the \eEDM~experiment on HfF$^+$ ions \cite{Cairncross:2017}. Trapped ions can be then laser-cooled to the Doppler limit. In this case the long lifetime becomes an advantage, since it allows to achieve lower temperatures \cite{Tarbutt:2018}. In order to estimate lifetime of the (1)3/2 state of AcOH$^+$, the dipole moment function was calculated for the (1)3/2-(1)1/2 transition (see Fig.~\ref{fig:tdm}). The lifetime was estimated using the approximate sum rule~\cite{Tellinghuisen:1984,Krumins:2020} and is expected to be $\tau \sim 110$ $\mu$s. This corresponds to the Doppler limit of order $T_D \sim 4$ nK. Such a long lifetime of the (1)3/2 state is not surprising, since this state actually corresponds to the $^2\Delta$ doublet (see Table~\ref{tab:equilibrium}) and is formally $E1$-forbidden in the absence of spin-orbit coupling.

\subsection{
Calculations of nuclear MQM
}
The magnetic quadrupole moment of a nucleus due to the electromagnetic current of a single nucleon with mass $m$ is defined by the second order tensor operator \cite{Sushkov:84},
\begin{align} \label{eq:MQMTensor}
\begin{split}
\hat{M}_{kn}^{\nu} = \dfrac{e}{2m}\left[3\mu_{\nu}\left(r_k\sigma_n + \sigma_kr_n - \dfrac{2}{3}\delta_{kn}\hat{\boldsymbol{\sigma}}\textbf{r}\right) \right. \\
\left. + 2q_{\nu}\left(r_kl_n + l_kr_n\right)\right]
\end{split}
\end{align}
where $\nu = p,n$ for protons and neutrons, respectively, and $m$, $\mu_{\nu}$ and $q_{\nu}$ are the mass, magnetic moment in nuclear magnetons $\dfrac{e}{2m}$ and charge of the nucleon, respectively.  In the case of axially symmetric nucleus  for an orbital with definite spin projection on the nuclear axis $\Sigma$ and orbital angular momentum projection $\Lambda$ we may give the following estimate for expectation value of MQM
\begin{align} \label{eq:Mzz}
M_{zz}^{\nu} = 4 \dfrac{e}{2m}\mu_z < r_z>, 
\end{align}
where $\mu_z= 2 \Sigma \mu_{\nu} + q_{\nu} \Lambda$ is the projection of the nucleon magnetic moment  and $< r_z>$ is the expectation value of the radius vector ${\bf r}$. The latter vanishes in the absence of the octupole deformation, therefore $< r_z> \sim \beta_3 R$, where $R\approx A^{1/3} 1.2$ fm is the nuclear radius.  

We presented MQM  in the intrinsic frame which rotates with the nucleus. We have to find MQM in  the  laboratory frame. 
 Similar to a polar diatomic molecule with a non-zero electron angular momentum, nucleus with an octupole deformation and non-zero nucleon angular momentum  has a doublet of close opposite parity rotational states $\ket{I^{\pm}}$ with the same angular momentum $I$ ($\ket{I^{\pm}}=\frac{1}{\sqrt{2}} (\ket{\Omega} \pm \ket{-\Omega})$, where $\Omega=\Sigma +\Lambda$ is  the projection of $I$ on to the nuclear axis). 
 In the case of ordinary electric quadrupole moment $Q$, which conserves $\mathcal{P}$ and $\mathcal{T}$ symmetries,  we have the relation $\braket{\Omega | Q_{zz} | \Omega}= \braket{ - \Omega | Q_{zz} | - \Omega}$ and the relation between the intrinsic value  $Q_{zz}$ and laboratory value $Q$ in the ground rotational state \cite{BohrMott}:
 
 \begin{align} \label{eq:RotationalFactor}
Q = \dfrac{I\left(2I - 1\right)}{\left(I + 1 \right)\left(2I + 3\right)}Q_{zz},
\end{align}
where $I=I_z= \left|\Omega\right|$ is the projection of total nuclear angular momentum (nuclear spin $I$) on the symmetry axis, $\Omega={\bf I \cdot n}$. This expression  for $Q$ shows that only in nuclei with spin $I > 1/2$ can we detect these second order tensor properties. In the case of MQM we have \\
 $\braket{\Omega | M_{zz} | \Omega}= - \braket{ - \Omega | M_{zz} | - \Omega}$ and  the laboratory value of $M$ vanishes in the states of definite parity  $\ket{ I^{\pm} }=\frac{1}{\sqrt{2}} (\ket{\Omega} \pm \ket{-\Omega})$ which have equal weights of $\Omega$ and   $-\Omega$ components. This is  a consequence  of $\mathcal{T}$ and $\mathcal{P}$ conservation. 

However, the states of  this doublet are mixed by $\mathcal{T},\mathcal{P}-$violating interaction $W$. The mixing coefficient is:
\begin{equation}\label{alpha}
 \alpha_{+-}=\frac{\braket{I^-| W| I^+}}{E_+  -  E_-} . 
\end{equation}
This mixing produces non-equal weights of $\Omega$  and  $-\Omega$ ,  $(1+  \alpha_{+-})^2/2$ and $(1 - \alpha_{+-})^2/2$, and leads to the non-zero expectation value of $<{\bf I \cdot n}>$, i.e.  the mixing  polarises the  nuclear axis ${\bf n}$ along the nuclear spin ${\bf I}$ \cite{Spevak:97,Auerbach:1996}:

\begin{equation}\label{n}
 <n_z>= 2 \alpha_{+-} \frac{I_z}{I+1}\,.
\end{equation}
As a result  all intrinsic $\mathcal{T}$,$\mathcal{P}$-odd nuclear moments show up in the laboratory frame. For MQM in the ground nuclear state using Eqs. (\ref{eq:Mzz}-\ref{eq:RotationalFactor}) we obtain
\begin{align} \label{eq:MQMRotationalFactor}
\nonumber
M = 2 \alpha_{+-} \dfrac{I\left(2I - 1\right)}{\left(I + 1 \right)\left(2I + 3\right)}  M_{zz}=\\
8 \dfrac{I\left(2I - 1\right)}{\left(I + 1 \right)\left(2I + 3\right)} \alpha_{+-}  \dfrac{e}{2m}\mu_z < r_z>\,.
\end{align}
  According to Ref. \cite{Spevak:97} the $\mathcal{T}$,$\mathcal{P}$-violating matrix element is approximately equal to
   \begin{equation}\label{W}
   \braket{I^-| W| I^+} \approx \frac{\beta_3 \eta}{A^{1/3}} [ \textrm{eV}].
  \end{equation}
  Here $\eta$ is the dimensionless strength constant of the nuclear $\mathcal{T},\mathcal{P}$-violating potential $W$:
   \begin{equation}\label{eta}
 W= \frac{G}{\sqrt{2}} \frac{\eta}{2m} ({\bf \sigma \nabla}) \rho ,
   \end{equation}
where $G$ is the Fermi constant, $m$ is the  nucleon mass and $\rho$ is the nuclear number density. Nuclear magnetic moment in the intrinsic frame is related  to $\mu$ in the laboratory frame 
\begin{equation}\label{mu}
 \mu=  \frac{I}{I+1}\mu_z\,.
\end{equation}
Using   $< r_z> \approx  1.2 \beta_3 A^{1/3} $ fm we obtain  nuclear MQM 
\begin{align} \label{eq:Mlab}
M\approx  \dfrac{2I - 1}{2I + 3} <(\beta_3)^2>  \frac{  \textrm{eV}} {E_+  -  E_-}  \mu \eta \, e \,\textrm{fm}^2\,.
\end{align}
Note that MQM in Eq. (\ref{eq:Mlab})  is quadratic in the octupole deformation parameter. Therefore, it is sufficient to have a soft octupole deformation mode, i. e. dynamical deformation,  which actually gives values of $ <(\beta_3)^2> $ comparable to that for the static octupole deformation. 

Within the meson exchange theory the $\pi$-meson exchange gives the dominating contribution to the $\mathcal{T}$,$\mathcal{P}$-violating nuclear forces
 \cite{Sushkov:84}. According to Ref. \cite{FDK14} the neutron and proton constants in the $\mathcal{T}$,$\mathcal{P}$-odd potential  (\ref{eta}) may be presented as  
\begin{align}
\eta_n=-\eta_p =   (-  g \bar{g}_{0} + 5 g \bar{g}_{1} + 2 g  \bar{g}_{2} ) 10^{6}, 
\end{align} 
 where $g$ is the strong $\pi$-meson-nucleon  interaction constant and ${\bar g}_0$, ${\bar g}_1$, ${\bar g}_2$ are the  $\pi$-meson-nucleon $\mathcal{CP}$-violating interaction  constants in the isotopic channels $T=0,1,2$.

We can express $\eta$ in terms of the QCD $\theta$-term constant. Using 
results presented in Refs. ~\cite{Vries:2015,Bsaisou:15}
\begin{align}
g \bar{g}_{0} &=-0.21  \, \bar{\theta}, \\
g \bar{g}_{1} &= 0.046  \,  \bar{\theta}, 
\end{align}
 we obtain  
\begin{align} \label{etawrttheta}
 \eta_n=-\eta_p =  4 \times 10^{5} \ \bar{\theta},\\ 
\end{align}
We also can express $\eta$ via the quark chromo-EDMs ${\tilde d_u}$ and  ${\tilde d_d}$. Using relations  $g {\bar g}_0 = 0.8 \times 10^{15}({\tilde d_u} +{\tilde d_d})$/cm,  $g {\bar g}_1 = 4 \times 10^{15}({\tilde d_u} - {\tilde d_d})$/cm~\cite{PR} we obtain:
\begin{align}
 \eta_n=-\eta_p =  ( 2 (\tilde{d_{u}} - \tilde{d_{d}}) - 0.1 (\tilde{d_{u}} + \tilde{d_{d}}) ) 10^{22}/\text{cm}.
\end{align}
In the expressions above interaction constants have opposite sign for valence proton and valence neutron (which in this case is unparied nucleon, which defines a non-zero value of $\Omega$). However,  magnetic moments also have opposite signs for proton and neutron,  so overall sign of the product  $\mu \eta$  for valence protons and neutrons is the same.  

The $^{227}$Ac nucleus  has a half-life of 21.8 years. $^{227}$Ac  is a product of uranium  decay chain. It is also produced in nuclear reactors by neutron capture by $^{226}$Ra. This isotope is commercially available. It has spin $I=3/2$ and magnetic moment $\mu=1.1(1)$.  The interval between opposite parity levels, which are mixed by the $\mathcal{T}$,$\mathcal{P}$-odd interaction, is $E(\frac{3}{2}^+) - E(\frac{3}{2}^-)=27.37$ keV.  
Experimental nuclear excitation spectra in this nucleus satisfy criteria for the octupole deformation. This nucleus has a proton above  $^{226}$Ra, which according to Ref.~\cite{Afanasjev2016} has an octupole deformation with  $\beta_3 =0.134 $. In the experimental paper Ref. \cite{Verstraelen:2019} it was found that   $\beta_3=0.1$  for $^{227}$Ac and $^{225}$Ac. Substitution of these parameters to formulas presented above give:
\begin{align} \label{MetaAc}
M(\eta) \approx 1. 3 \cdot 10^{-33} \eta_p \, e \, \textrm{ cm}^2,
\end{align}
\begin{align} \label{MgAc}
M(g) \approx 1.3 \cdot 10^{-27} 
(-  g \bar{g}_{0} + 5 g \bar{g}_{1} + 2 g  \bar{g}_{2} ) \, e \,  \textrm{ cm}^2,
\end{align}
\begin{align} \label{MthetaAc}
M(\theta) \approx 5  \cdot 10^{-28}  \bar{\theta} \, e\, \textrm{ cm}^2,
\end{align}
\begin{align} \label{MdAc}
M(d) \approx  1.3 \cdot 10^{-11} ( 2  (\tilde{d_{u}} - \tilde{d_{d}})-0.1 (\tilde{d_{u}} + \tilde{d_{d}}) ) \, e \,  \textrm{ cm}.
\end{align}
Energy interval within the doublet  $E_+  -  E_-$=27 keV is  much smaller than the interval between the opposite parity orbitals in spherical nuclei ( $\sim$ 8 MeV). Therefore, value of MQM is  1 -- 2 orders of magnitude bigger than MQM of a spherical nucleus estimated in Ref. \cite{Sushkov:84}. 

The $^{225}$Ac nucleus  has a half-life of 10 days. A  similar calculations show that the results for $^{225}$Ac is 1.5 times smaller since the energy interval is bigger, $E(\frac{3}{2}^+) - E(\frac{3}{2}^-)=  40.1$ keV.   

Note that a comparable contribution to MQM of actinium isotopes is given by a spin hedgehog mechanism which requires quadrupole deformation (but does not require octupole deformation) \cite{F94}.
Magnetic quadrupoles in this case have collective nature, somewhat similar to collective electric quadrupoles in deformed nuclei.  Actinium isotopes have a significant quadrupole deformation and posses collective MQM with the values close  to that of  $^{229}$Th (neighbour of Ac in periodic table) estimated in Refs. \cite{FDK14,Lackenby:2018}.

\subsection{Relevance to $\mathcal{T},\mathcal{P}$-violation search experiments}

\begin{table}[]
\caption{Calculated values of the effective electric field (\Eeff), parameter of the scalar-pseudoscalar nucleus-electron interaction ($W_{T,P}$).}
\label{TPResults}
\begin{tabular}{lrrr}
\hline
\hline
Method     & \Eeff,     & ~~~$W_{T,P}$,  & ~~~$W_M$\\
           &  GV/cm     & kHz            & ~~~10$^{33} \frac{\textrm{Hz}}{e\cdot \textrm{cm}^2}$\\
\hline
DHF, DC            & -47.6         & -131.3   & -1.325     \\
97e-CCSD, DC       & -58.7         & -161.2   & -1.744     \\
\\
97e-CCSD(T), DC    & -57.4         & -157.6   & -1.701     \\
Basis correction   & 1.4           & 3.6      &  0.027     \\
Gaunt              & 1.0           & 2.0      &  0.024     \\
Final              & -55.0         & -152.1   & -1.651     \\   
\hline
\hline
\end{tabular}
\end{table}

As it was mentioned above, the $l$-doubling leads to the existence of the two closely spaced levels of opposite parity. The strength of the external electric field required to polarize the molecule is proportional to the energy spacing. Levels with $l=2$ possess very small spacing, but can hardly be populated in actual experiment, so only the $l=1$ levels are to be considered as the working ones. To the lowest order the energy spacing $\Delta E_J$ for the $l=1$ states can be roughly estimated using the formula \cite{HerzbergVol3}:
\begin{equation}
    \Delta E_J \approx \frac{B^2}{\nu_2} (v_2 + 1) J (J+1),
\end{equation}
where $B$ is the rotational constant, $\nu_2$ -- frequency of the bending mode, $v_2$ -- vibrational quantum number for the bending mode. For the (0,$1^1$,0) $J=1$ state of AcOH$^+$ this estimate gives $\Delta E_{J=1} \approx 15$ MHz, which is very close to the corresponding spacing predicted for RaOH ($\approx$14.5~MHz~\cite{Zakharova:RaOH:2021}) and nearly two times less than the spacing in YbOH ($\approx$26~MHz~\cite{Zakharova:2021}), as expected from the smaller $v_2$ frequency in the latter case. Thus the external electric field strength required to polarize the system is expected to be of the same order as for YbOH ($\sim$ 100 V/cm), and even smaller.

To estimate the lifetime of the excited (0,1${}^1$,0) vibrational state, we have approximated the permanent dipole moment of AcOH$^+$ as a linear function and calculated its slope at the DFT level of theory (see the Computational details section). The Wigner-Eckart conditions were additionally imposed to approximately get rid of the origin dependence of the dipole moment of a charged system. Our rough estimation gives transition moment $|d| \approx 0.13$ a.u. Combining this value with the predicted excitation energy $\nu_2 = 423$ cm$^{-1}$, we arrive at the lifetime of $\approx 0.4$ s, which is comparable to the lifetime of the ${}^3\Delta_1$ state in HfF${}^+$ \cite{Cornell:2017}.

Table~\ref{TPResults} gives calculated values of the molecular constants of $\mathcal{T}, \mathcal{P}$-violation interactions for the ground electronic state of AcOH$^+$. Perturbative triple cluster amplitudes contribute about 2.5\% to the final values. According to Refs.~\cite{Skripnikov:15a,Skripnikov:15b} one can expect that higher-order correlation contributions are small compared to this value. One can also see from Table~\ref{TPResults} that the basis set size extension corrections are rather small, proving a good convergence. This is also true for the Gaunt contribution estimation. Note also, that when the Breit interaction is considered, one should include additional terms in the \eEDM~interaction Hamiltonian~(\ref{Wd2}), see Refs.~\cite{Lindroth:89} for details and Ref.~\cite{Skripnikov:16b} for discussion. Therefore, the given ``Gaunt'' correction can be considered only as rough estimate for the case of \Eeff. The given values of $W_M$, \Eeff, $W_{T,P}$ have been calculated for the theoretical equilibrium geometry of AcOH$^+$ determined in the present paper (see above). In principle, it is possible to calculate the values, averaged over the vibrational wavefunction. In Refs.~\cite{Skripnikov:15a,Skripnikov:15b} it has been shown that this contribution is small for diatomic molecules. Recently, it has been also shown \cite{Zakharova:2021} that for YbOH the vibrationally averaged value of the effective electric field is very close to the value, calculated for the equilibrium geometry. Therefore, we neglect this contribution in the present study. Taking into account the data in Table~\ref{TPResults} as well as our previous experience~\cite{Skripnikov:16b,Skripnikov:17c} one estimate the uncertainty of calculated molecular constants of about 4\%. 

The resulting value of $W_M \Omega$ for AcOH$^+$ is $\approx$1.5 times bigger than that of YbOH~\cite{Maison:2019b,denis2019enhancement} and is slightly smaller than in ThO~\cite{Skripnikov:14a,Skripnikov:14aa}, but twice as large as for HfF$^+$~\cite{Skripnikov:17b} (see also Table II Ref.~\cite{Maison:20a} for comparison).

As it is mentioned above, there are several isotopes of Ac suitable for experiments to search for the nuclear MQM. According to Eq.~(\ref{MQMshift}) the $W_M$ constant is required to extract the value of the nuclear MQM from the experimental energy shift. Eqs.~(\ref{MetaAc})--(\ref{MdAc}) can be used further to extract constraints on such parameters as the quantum chromodynamics vacuum angle $\bar{\theta}$ and quark chromo-EDMs.
Note, that the characteristic energy shift induced by MQM is proportional to $W_M M$. This product expressed in terms of $\bar{\theta}$ is nearly the same as in the $^{173}$YbOH~\cite{Maison:2019b,denis2020enhanced} molecule (see also Table II Ref.~\cite{Maison:20a} for comparison with other molecules). Both AcOH$^+$ and YbOH molecules can be used to search for other $\mathcal{T}$,$\mathcal{P}$-violating effects as the electron EDM and scalar-pseudoscalar nucleus-electron interaction. Therefore, in order to perform more clear interpretation one has to have several independent measurements with different sensitivities to different $\mathcal{T}$,$\mathcal{P}$-violating sources. Here we can also note, that due to different number of protons in Ac and Yb the sensitivities of AcOH$^+$ and YbOH to \eEDM~ and $k_{T,P}$ are different.

\section{Conclusion}

We have considered the AcOH${}^+$ molecular ion viable for measuring $\mathcal{T,P}$-violation effects. To our best knowledge, AcOH${}^+$ is the first polyatomic molecular ion which is predicted to be laser coolable. The wide variety of relatively long-lived and experimentally available isotopes allows one to use this molecular ion to perform measurements of fundamentally different effects, e.g. electron EDM and nuclear magnetic quadrupole moment. The estimates for the nuclear MQM based on nuclear structure calculations can be used for interpretation of the experimental data in terms of the fundamental QCD parameters. Finally, based on the analogy with the YbOH/YbOCH${}_3$ molecules, one can expect that the closed optical cycle can also exist for AcOCH$_3^+$. If true, this will reveal the path to the new generation of $\mathcal{T,P}$-violation search experiments with polyatomic chiral molecular ions.

\section{Acknowledgements}

We are grateful to Anna Zakharova for useful discussions. Electronic structure calculations have been carried out using computing resources of the federal collective usage center Complex for Simulation and Data Processing for Mega-science Facilities at National Research Centre ``Kurchatov Institute'', http://ckp.nrcki.ru/.

The computational study of AcOH$^+$ excited states and transition moments performed at NRC ``Kurcgatov Institute'' -- PNPI was supported by the Russian Science Foundation (Grant No. 20-13-00225). Coupled cluster calculations of the $\mathcal{T},\mathcal{P}$-violating electronic structure parameters performed at NRC ``Kurchatov Institute'' -- PNPI were supported by the Russian Science Foundation (Grant No. 18-12-00227). Calculations of the matrix elements of the effective electric field performed at SPbSU were supported by the foundation for the advancement of theoretical physics and mathematics ``BASIS'' grant according to Project No. 21-1-2-47-1. V.F. is supported by the Australian Research Council grants DP190100974 and  DP200100150, and the JGU Gutenberg Fellowship for the nuclear calculations.

\bibliographystyle{apsrev}

\end{document}